\documentclass[12pt]{iopart}
\usepackage{epsfig,amssymb,psfig}

\def\spose#1{\hbox to 0pt{#1\hss}}
\def\lesssim{\mathrel{\spose{\lower 3pt\hbox{$\mathchar"218$}}
 \raise 2.0pt\hbox{$\mathchar"13C$}}}
\def\gtrsim{\mathrel{\spose{\lower 3pt\hbox{$\mathchar"218$}}
 \raise 2.0pt\hbox{$\mathchar"13E$}}}

\def\<{\langle}
\def\>{\rangle}
\newcommand\onlinecite{\cite}

\begin{document}

\title{ 
The critical behaviour of $3D$ Ising spin glass models:
universality and scaling corrections
} 

\author{Martin Hasenbusch,$^{1}$
Andrea Pelissetto,$^2$
and Ettore Vicari$\,^3$ }

\address{$^1$
Institut f\"ur Theoretische Physik, Universit\"at Leipzig,\\
Postfach 100 920, D-04009 Leipzig, Germany}

\address{$^2$
Dip. Fisica dell'Universit\`a di Roma ``La Sapienza" and INFN, \\
I-00185 Roma, Italy}

\address{$^3$ Dip. Fisica dell'Universit\`a di Pisa and
INFN, I-56127 Pisa, Italy}

\ead{
Martin.Hasenbusch@df.unipi.it,
Andrea.Pelissetto@roma1.infn.it,
Ettore.Vicari@df.unipi.it}

\date{\today}

\begin{abstract}
  We perform high-statistics Monte Carlo simulations of three
  three-dimensional Ising spin glass models: the $\pm J$ Ising model for two
  values of the disorder parameter $p$, $p=1/2$ and $p=0.7$, and the
  bond-diluted $\pm J$ model for bond-occupation probability $p_b = 0.45$. A
  finite-size scaling analysis of the quartic cumulants at the critical point
  shows conclusively that these models belong to the same universality class
  and allows us to estimate the scaling-correction exponent $\omega$ related
  to the leading irrelevant operator, $\omega=1.0(1)$. We also determine
  the critical exponents $\nu$ and $\eta$. Taking into account the scaling
  corrections, we obtain $\nu=2.53(8)$ and $\eta=-0.384(9)$.
\end{abstract}


\maketitle




The most peculiar aspect of critical phenomena is the universality of the
asymptotic behaviour in a neighborhood of the critical point.  In experiments
and Monte Carlo (MC) simulations the possibility of approaching the critical
point (and/or the infinite-volume limit) is generally limited.  Therefore, an
accurate determination of the universal critical behaviour requires a good
control of the scaling corrections.  This is particularly important for
systems with disorder and frustration, such as spin glasses, where severe
technical difficulties make it necessary to work with systems of relatively
small size.  Even though the critical behaviour of Ising spin glass models has
been much investigated numerically in the last two decades, see, e.g.,
\cite{KKY-06,CHT-06,Jorg-06,PPV-06,PRT-06,PC-05,NEY-03,MC-02,
BCFMPRTTUU-00,PC-99,MC-99,MPR-98},
it is not yet clear how reliable the numerical results are.  For
instance, the estimates of the critical exponents have significantly
changed during the years, as shown by the results reported in
\onlinecite{KKY-06}. Moreover, the most recent MC studies, see,
e.g., \onlinecite{KKY-06,CHT-06}, find significant discrepancies
among the estimates of the correlation-length exponent $\nu$ obtained
from the finite-size scaling (FSS) at $T_c$ of different observables,
such as the temperature derivatives of $\xi/L$, of the Binder
cumulant, and of the susceptibility. For instance, for the
bimodal Ising model \onlinecite{KKY-06} quotes $\nu=2.39(5)$,
$\nu=2.79(11)$, and $\nu=1.527(8)$, from the analysis of these three
quantities.  These differences indicate the presence of sizeable
scaling corrections. However, the data are not precise enough to allow
for scaling corrections in the analyses.  In order to reduce their
effects, \onlinecite{CHT-06} has proposed an alternative purely {\em
phenomenological} scaling form inspired by the high-temperature
behavior, which affects only the {\em analytic} scaling corrections
and apparently reduces the differences among the estimates of $\nu$.
Some attempts to determine the {\em nonanalytic} scaling corrections
have been reported in
\onlinecite{PC-99,BCFMPRTTUU-00,MC-02,PPV-06}, but results are
quite imprecise.

Summarizing, little is known about the scaling corrections in Ising
spin glass models, even though it is now clear that their
understanding is crucial for an accurate determination of the critical
behavior. Here we report a MC study, which represents a substantial
progress in this direction. Indeed, by a FSS analysis of
renormalisation-group (RG) invariant quantities we are able to obtain
a robust estimate of the leading scaling-correction exponent
$\omega$. This allows us to analyze the MC data for the critical
exponents taking the scaling corrections into account.  Estimates
obtained from different observables are now in agreement.

We consider the $\pm J$ Ising model on a cubic lattice of linear size
$L$ with periodic boundary conditions. The corresponding Hamiltonian
is
\begin{equation}
H = - \sum_{\langle xy \rangle} J_{xy} \sigma_x \sigma_y,
\label{lH}
\end{equation}
where $\sigma_x=\pm 1$, the sum is over pairs of nearest-neighbor lattice sites,
and the exchange interactions $J_{xy}$ are uncorrelated quenched random
variables with probability distribution $P(J_{xy}) = p \delta(J_{xy} - 1) +
(1-p) \delta(J_{xy} + 1)$.  The usual bimodal Ising spin glass
model~\cite{EA-75}, for which $[J_{xy}]=0$ (brackets indicate the average over
the disorder distribution), corresponds to $p=1/2$.  For $p\neq 1/2$,
$[J_{xy}]=2p-1\neq 0$ and ferromagnetic (or antiferromagnetic)
configurations are energetically favored. A reasonable hypothesis is that,
along the transition line separating the paramagnetic and the spin glass
phase, the critical behaviour is independent of $p$, i.e., a nonzero value of
$[J_{xy}]$ is irrelevant, as found in mean-field models~\cite{Toulouse-80}.
The paramagnetic-glass transition line
extends for $p^*>p > 1-p^*$, where~\cite{HPPV-07-mgp} $p^*=0.76820(4)$.
For $1>p>p^*$
the low-temperature phase is ferromagnetic, and the transition belongs to the
randomly-dilute Ising universality class~\cite{HPPV-07-pmj}. We also consider
a bond-diluted $\pm J$ Ising model with bond-occupation probability
$p_b=0.45$, and equal probability for the values $\pm J$.

We focus on the critical behaviour of the overlap parameter $q_x \equiv
\sigma_x^{(1)} \sigma_x^{(2)}$, where $\sigma_x^{(i)}$ are independent
replicas with the same disorder $J_{xy}$.  If $G(x) \equiv [ \langle q_0 q_x
\rangle ]$, we define the susceptibility $\chi\equiv \sum_{x} G(x)$ and the
second-moment correlation length $\xi$
\begin{eqnarray}
\xi^2 \equiv  {1\over 4 \sin^2 (q_{\rm min}/2)} 
{\widetilde{G}(0) - \widetilde{G}(q)\over \widetilde{G}(q)},
\label{xidefffxy}
\end{eqnarray}
where $q = (q_{\rm min},0,0)$, $q_{\rm min} \equiv 2 \pi/L$,
and $\widetilde{G}(q)$ is the Fourier transform of $G(x)$.  We
also define 
\begin{eqnarray}
R_\xi\equiv\xi/L,\quad
U_{4}  \equiv { [ \mu_4 ]\over [\mu_2]^{2}}, \quad
U_{22} \equiv  {[ \mu_2^2 ]-[\mu_2]^2 \over [\mu_2]^2},
\label{Rdef}
\end{eqnarray}
where $\mu_{k} \equiv \langle \; ( \sum_x q_x\; )^k \rangle$.  The quantities
(\ref{Rdef}) are RG invariant. We call them phenomenological couplings and
denote them by $R$ in the following.

Let us first summarize some basic results concerning FSS, which allow us to
understand the role of the {\em analytic} and {\em nonanalytic} scaling
corrections.  We consider two Ising spin glass systems coupled by 
an interaction $h\sum_x q_x = h \sum_x \sigma^{(1)}_x \sigma^{(2)}_x$,
in a finite volume of linear size $L$. 
The singular part of the corresponding disorder-averaged free
energy density ${\cal F}$, which encodes the critical behavior, behaves as
\begin{eqnarray}
&&{\cal F}_{\rm sing}(\beta,h,L) =
 L^{-d} F( u_h L^{y_h}, u_t L^{y_t}, \{ v_i L^{y_i}\}) \label{FscalL}\\
&& = L^{-d} f(u_h L^{y_h}, u_t L^{y_t}) +
  v_\omega L^{-d-\omega} f_{\omega}(u_h L^{y_h}, u_t L^{y_t})  + \ldots,
\nonumber
\end{eqnarray}
where $u_h$ and $u_t$ are the scaling fields associated respectively with $h$
and $t\sim T-T_c$ (their RG dimensions are $y_h=(d+2-\eta)/2$ and
$y_t=1/\nu$), and $v_i$ are irrelevant scaling fields with $y_i<0$.  The
leading {\em nonanalytic} correction-to-scaling exponent $\omega$ is related
to the RG dimension $y_\omega$ of the leading irrelevant scaling field
$v_\omega\equiv v_1$, $\omega= - y_\omega$.  The scaling fields are analytic
functions of the system parameters---in particular, of $h$ and $t$---and are
expected not to depend on $L$. Note also that the size $L$ is expected to be
an exact scaling field for periodic boundary conditions. For a general
discussion of these issues, see \onlinecite{SS-00,HPPV-07} and references
therein.  In general, $u_t$ and $u_h$ can be expanded as
\begin{eqnarray}
&&u_h = h \bar{u}_h(t) + O(h^3),\quad \bar{u}_h(t) = a_h + a_1 t + O(t^2),\\
&&u_t = c_t t + c_{02} t^2 + c_{20} h^2 + c_{21} h^2 t + O(t^3,h^4,h^4 t)  ,
\nonumber
\end{eqnarray}
where we used the fact that the free energy is symmetric under $h\to -h$. In
the expansion of $u_{h,t}$ around the critical point $h,t=0$, the terms beyond
the leading ones give rise to {\em analytic} scaling corrections. There are
also analytic corrections due to the regular part of the free energy; 
since they scale as $L^{\eta - 2} \sim L^{-2.4}$, 
they are negligible in the present case. The scaling behaviour of zero-momentum
thermodynamic quantities can be obtained by performing appropriate derivatives
of ${\cal F}_{\rm sing}$ with respect to $h$.  For instance, the overlap
susceptibility $\chi = \partial^2 {\cal F}/ \partial h^2 |_{h=0}$ behaves as
\begin{equation}
\chi = 
L^{2-\eta} \bar{u}_h(t)^2 g(u_t L^{y_t}) 
+ L^{2-\eta-\omega} g_\omega(u_t L^{y_t}) + \ldots
\label{chiexp_1}
\end{equation}
The FSS of the phenomenological couplings is given by
\begin{eqnarray}
R(\beta,L) &=&  r(u_t L^{y_t}) + r_\omega(u_t L^{y_t}) L^{-\omega}  + \ldots
\nonumber \\
&=& 
R^* + r'(0) c_t \, t L^{y_t}   + \ldots +  c_\omega \, L^{-\omega} + \ldots,
\label{Rexp_1}
\end{eqnarray}
where $R^*\equiv r_0(0)$ and $c_\omega = r_\omega(0)$. 
In the case of $U_4$, this behaviour can be proved by taking the appropriate 
derivatives of ${\cal F}$ with respect to $h$. A similar discussion applies
to $U_{22}$ and $\xi/L$, see Sec. 3.1 of \cite{HPPV-07} for details. 
The exponent $\nu$ can
be computed from the FSS of the derivative $R'\equiv dR/d\beta$ at $\beta_c$,
or from that of the ratio $\chi'/\chi$, where $\chi'\equiv d\chi/d\beta$.  At
$T=T_c$, setting $t=u_t=0$ in the above-reported equations, we obtain:
\begin{eqnarray}
&& R = R^* + c_1 L^{-\omega} + \ldots, \label{rin}\\
&& \chi = c L^{2-\eta} ( 1 + c_1 L^{-\omega} + \ldots), \label{chibc} \\
&& R' = c L^{1/\nu} ( 1  + c_1 L^{-\omega} + \ldots), \label{Rpbc} \\
&& \chi' =  c L^{2-\eta+1/\nu} 
( 1  + c_1 L^{-\omega} + \ldots + a_1 L^{-1/ \nu} + \ldots) .
\label{chipbc}
\end{eqnarray}
Note that, unlike the temperature derivative $R'$ of an RG invariant quantity,
$\chi'$ also presents an $L^{-1/\nu}$ scaling correction, due to the analytic
dependence on $t$ of the scaling field $u_h$ (for this reason we call it {\em
  analytic} correction).  Since, as we shall see, $1/\nu\approx 0.4$ and
$\omega\approx 1.0$, the scaling corrections in $\chi'$ decay slowlier than
those occurring in $R'$. This makes $\chi'/\chi$ unsuitable for a precise
determination of $\nu$ and explains the significant discrepancies observed in
\onlinecite{KKY-06}.

Instead of computing the various quantities at fixed Hamiltonian parameters,
we consider the FSS keeping a phenomenological coupling $R$ fixed at a given
value $R_{f}$ \cite{Has-99,HPPV-07}.  This means that, for each $L$, we
determine $\beta_f(L)$, such that $R(\beta=\beta_f(L),L) = R_{f}$, and then
consider any quantity at $\beta = \beta_f(L)$.  The value $R_{f}$ can be
specified at will, as long as $R_f$ is taken between the high- and
low-temperature fixed-point values of $R$.  For generic values of $R_f$,
$\beta_f$ converges to $\beta_c$ as $\beta_f-\beta_c=O(L^{-1/\nu})$, while at
$R_{f} = R^*$, cf. (\ref{rin}), $\beta_f-\beta_c=O(L^{-1/\nu-\omega})$.
One can easily show that the FSS behaviour at fixed $R_f=R^*$ is given by the
same general formulas derived at $T_c$.  In the case of another
phenomenological coupling $R_\alpha$ we have
\begin{equation}
\bar{R}_\alpha(L)\equiv R_\alpha[\beta_f(L),L]\approx
\bar{R}_\alpha^* + c_\alpha L^{-\omega} + \ldots ,
\label{barr}
\end{equation}
where $\bar{R}_\alpha^*$ is universal but depends on $R_f$.  If $R_f$ differs
from $R^*$, $\chi$ and $R'$ (but not the phenomenological couplings
$R$) also present $L^{-k/\nu}$ corrections with amplitudes proportional to
$(R_f - R^*)^k$.

\begin{table}
\caption{\sl \label{MCinfo}
Parameters for the simulations of the $\pm J$ model at $p=0.5$
for $L\ge 12$.
We use the random-exchange method with $N_{\beta}$ temperatures
between $\beta_{\rm min}$ and $\beta_{\rm max} = 0.895$. 
An elementary iteration is composed by $n_{\rm met}$ Metropolis 
full sweeps of all $N_\beta$ systems
followed by a temperature-exchange attempt of all 
pairs corresponding to nearby temperatures.
The length of the each run is $48 n_{\rm iter}$ iterations; 
the first $20 n_{\rm iter}$ are discarded for equilibration.
The total number of Metropolis sweeps per sample and $\beta$ is 
$n_{\rm tot} = 48 n_{\rm iter} n_{\rm met}$.
The CPU time refers to a single core of a dual-core 2.4 GHz 
AMD Opteron processor.
}
\begin{center}
\begin{tabular}{rrrrrrlc}
\hline
 $L$ &  samples/64 & $n_{\rm met}$ & $n_{\rm iter}$ & 
   $n_{\rm tot}/10^3$ & $N_{\beta}$ & $\beta_{\rm min}$ &
   CPU time in days \\
\hline
12  &  106812 &  10 & 400  & 192  & 10   & 0.54& 308 \\
13  &   38282 &  10 & 600  & 288  & 10   & 0.54& 210 \\
14  &   31600 &  50 & 200  & 480  & 10   & 0.62& 361 \\
16  &   24331 &  10 & 1000 & 480  & 20   & 0.52& 831 \\
20  &    1542 &  20 & 2000 & 1920 & 32   & 0.5125 & 658  \\
24  &     717 &  25 & 2500 & 3000 & 32   & 0.5125 & 826  \\
28  &     285 &  60 & 2500 & 7200 & 20   & 0.6575 & 782  \\
\hline
\end{tabular}
\end{center}
\end{table}

In the MC simulations we employ the Metropolis algorithm, the random-exchange
method~\cite{raex}, and multispin coding.  We simulate the $\pm J$ Ising model
at $p=0.5$ for $L$=3-14,16,20,24,28, at $p=0.7$ for $L$=3-12,14,16,20, and the
bond-diluted model at $p_b=0.45$ for $L$=4-12,14,16.  We average over a large
number $N_s$ of disorder samples: $N_s\approx 6.4\cdot 10^6$ up to $L=12$,
$N_s/10^3\approx 2400,2000,1500,100,46,18$, respectively for
$L=13,14,16,20,24,28$ in the case of the $\pm J$ Ising model at $p=0.5$.
See Table~\ref{MCinfo} for details.
Similar statistics are collected at $p=0.7$, while for the bond-diluted model
statistics are smaller (typically, by a factor of two for the small lattices
and by a factor of 6 for the largest ones). For each $L$ and model we perform
runs up to values of $\beta$ such that $R_\xi(\beta,L)$ is approximately 0.63,
which is close to the estimates of $R_\xi^*$ of \onlinecite{KKY-06}:
0.627(4) and 0.635(9) for an Ising model with bimodal and Gaussian distributed
couplings, respectively. We carefully check thermalization by using the
recipe outlined in \onlinecite{KKY-06}.  Estimates of the different
observables for generic values of $\beta$ close to $\beta_c$ 
are computed by using their second-order Taylor expansion around 
$\beta_{\rm max} = 0.895$.
We check the correctness of these estimates 
by comparing them with those obtained 
by using the Taylor expansion around the value of $\beta$
used in the random-exchange simulation that is closest to $\beta_{\rm max}$.
In total, the MC simulations took approximately 30 years of CPU-time on a 
single core of a 2.4 GHz AMD Opteron processor.

\begin{figure}[tb]
\centerline{\psfig{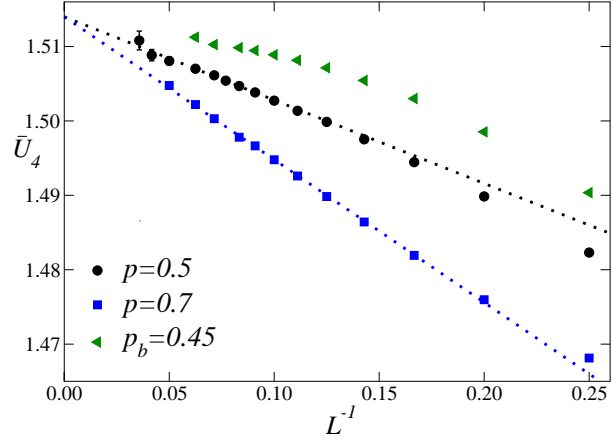}}
\vspace{2mm}
\caption{
Phenomenological coupling $\bar{U}_4(L)$ vs $L^{-1}$.
The lines are drawn to guide the eye.
}
\label{u4}
\end{figure}

We first perform a FSS analysis at fixed $R_\xi=0.63$. For sufficiently large
$L$, the FSS behaviour of $\bar{U}_4$ and $\bar{U}_{22}$ is given by
(\ref{barr}).  The MC estimates of $\bar{U}_4(L)$ are shown in
Fig.~\ref{u4} versus $1/L$. The results for the $\pm J$ Ising model at $p=0.5$
and $p=0.7$ fall quite nicely on two straight lines approaching the same point
as $L\to\infty$, indicating that $\omega\approx 1.0$.  In the case of the
diluted model the approach to the large-$L$ limit is faster: fits give
$\bar{U}_4(L) = \bar{U}^*_4 + e L^{-\epsilon}$ with $\epsilon\approx 2$.  This
indicates that $c_4\approx 0$ [see (\ref{barr})].  According to the RG,
this implies that the leading nonanalytic scaling correction 
is suppressed in any quantity.
Thus, the approach to the critical limit should be faster, as
already noted in \cite{Jorg-06}.  We fit the data to $\bar{U}^* + c
L^{-\epsilon}$, taking $\epsilon$ as a free parameter.  Using data for $L\ge
L_{\rm min}=8$, we obtain $\bar{U}_4^*=1.514(1)$, $\bar{U}_4^*=1.514(2)$, and
$\bar{U}_4^*=1.513(1)$ for the $\pm J$ model at $p=0.5$ and $p=0.7$, and the
bond-diluted model, respectively.  The fits of $\bar{U}_{22}$ to
$\bar{U}^*_{22} + c L^{-\epsilon}$ ($L_{\rm min} = 6$) give
$\bar{U}_{22}^*=0.1477(3)$, $\bar{U}_{22}^*=0.1481(8)$, and
$\bar{U}_{22}^*=0.1479(6)$, respectively for the $\pm J$ Ising model at
$p=0.5$ and $p=0.7$, and the bond-diluted model.  These results represent a
very accurate check of universality.

\begin{figure}[tb]
\centerline{\psfig{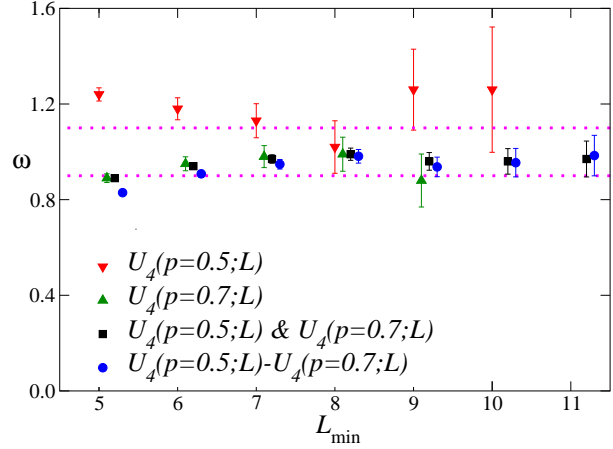}}
\vspace{2mm}
\caption{
Estimates of the leading scaling-correction exponent $\omega$.
For each $L_{\rm min}$, they are obtained by fitting only the data satisfying
$L\ge L_{\rm min}$.
The dotted lines correspond to the final estimate $\omega=1.0(1)$.
}
\label{omega}
\end{figure}

The analyses of $\bar{U}_4$ and $\bar{U}_{22}$ give also estimates of
$\omega$. The most precise ones are obtained from the analysis of $\bar{U}_4$.
In Fig.~\ref{omega} we show the results for $\omega$ as
obtained from fits of $\bar{U}_4$ to $\bar{U}^*_4 + c_p L^{-\epsilon}$ and of
fits of the difference $\bar{U}_4(p=0.5;L)-\bar{U}_4(p=0.7;L)$ to
$bL^{-\epsilon}$. To verify the stability of the results, we have repeated the
fits several times, each time including only the data satisfying 
$L\ge L_{\rm min}$. We estimate $\omega=1.0(1)$.  
As a check, we verify that the ratio $c_{22}/c_4$ is universal [$c_\#$ is the
scaling-correction amplitude appearing in (\ref{barr})], as predicted by
standard RG arguments.  Fits of $\bar{U}(L) - \bar{U}^*$ to $c L^{-\epsilon}$,
fixing $\bar{U}_4^*=1.514$, $\bar{U}_{22}^*=0.148$, $\epsilon=1.0$ ( $L_{\rm
  min} = 12$) give $c_{22}/c_4=0.19$ and $c_{22}/c_4=0.20$, respectively for
$p=0.5$ and $p=0.7$, which are in good agreement.

\begin{figure}[tb]
\centerline{\psfig{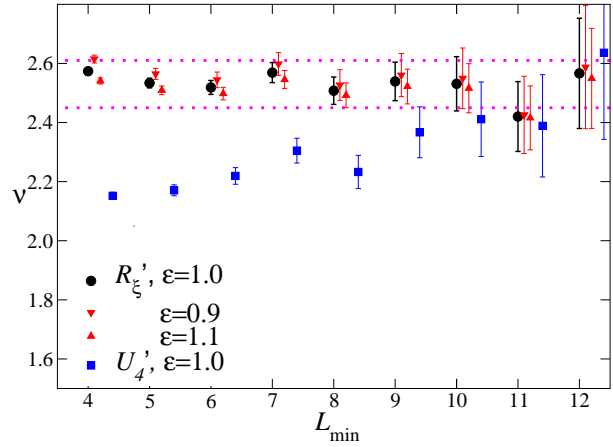}}
\vspace{2mm}
\caption{
Estimates of the exponent $\nu$ from the FSS analysis of
$R_\xi'$ and $U_4'$, for the $\pm J$ Ising model at $p=0.5$,
obtained by fitting the data to (\ref{ansatznu}).
The dotted lines correspond to the final estimate $\nu=2.53(8)$.
}
\label{nup0p5}
\end{figure}

Equation (\ref{barr}) holds for any chosen value of $R_f$. On the other hand,
$\chi$ and $R'$ do not present $O(L^{-1/\nu})$ corrections only if $R_f=R^*$.
Thus, before computing the critical exponents, we refined the estimate of
$R_\xi^*$ by performing a standard FSS analysis of $R_\xi$ which takes into
account the scaling corrections. Fixing $\omega = 1.0(1)$, we obtained
$R_\xi^* = 0.654(7)$, which is slightly larger than the estimates
reported in \onlinecite{KKY-06,Jorg-06}. For the $\pm J$ model at $p=0.5$
we also obtained $\beta_c=0.908(4)$. Then, in order to determine
the critical exponent $\nu$, we computed $R_\xi'$ and $U_4'$ at fixed
$R_{\xi,f}=0.654$. 
In Fig.~\ref{nup0p5} we show
results for the $\pm J$ Ising model at $p=0.5$, obtained by fitting $R_\xi'$
to
\begin{equation}
{\rm ln} R' = a + {1\over \nu} {\rm ln} L + b L^{-\epsilon},
\label{ansatznu}
\end{equation}
with $\epsilon=\omega=1.0(1)$, as a function of $L_{\rm min}$. They are quite
stable and lead to the estimate
\begin{equation}
\nu=2.53(6)[2],
\label{nuest}
\end{equation}
where the error in brackets takes into account the uncertainty on 
$\omega$.  Since $R_{\xi,f}=0.654$ is only
approximately equal to $R_\xi^*$, we may have residual $L^{-1/\nu}$
corrections.  The comparison with the same analysis at fixed
$R_\xi=0.63$ shows that their effect is negligible.
The results from fits of $U_4'$ to (\ref{ansatznu}),
shown in Fig.~\ref{nup0p5}, are substantially consistent. For
example, we find $\nu=2.41(13)$ for $\epsilon=1.0$ and $L_{\rm min} = 10$.
For the model at $p=0.7$ the fit of $R_\xi'$ ($L_{\rm min} = 8$) gives
$\nu=2.54(6)$.  Finally, we consider the bond-diluted model. If we use
$\epsilon=2$ (this is the value determined from the analysis of $\bar{U}_4$
and $\bar{U}_{22}$) we obtain $\nu=2.55(6)$ ($L_{\rm min} = 8$).  These
results are in good agreement with the estimate (\ref{nuest}).  

We estimate
the exponent $\eta$ by analyzing the susceptibility $\chi$ at fixed
$R_\xi=0.654$.  We fit ${\rm ln} \chi$ to $a + (2-\eta) {\rm ln} L +
b L^{-\epsilon}$.  In
the case of the $\pm J$ model at $p=0.5$, fixing $\epsilon=1.0$, we obtain
$\eta=-0.384(1),-0.384(4)$, for $L_{\rm min}=7,10$, respectively, with
$\chi^2/{\rm dof}\lesssim 1$.  Our final estimate is
\begin{equation}
\eta = -0.384(4)[0]\{5\},
\label{etaest}
\end{equation}
where the error in brackets is related to the uncertainty on $\omega$ and that
in braces gives the variation of the estimate as $R_{\xi,f}$ varies within two
error bars of $R_\xi^*=0.654(7)$. The other models give consistent, though less
precise results.  Finally, we have checked the scaling behaviour of $\chi'$,
which shows $L^{-1/\nu}$ scaling corrections for any value of $R_{\xi,f}$, see
(\ref{chipbc}). If we take them into account, the asymptotic behaviour of
$\chi'$ is consistent with the estimates of $\nu$ and $\eta$ obtained from
$R'$ and $\chi$. For instance, a fit of $\ln \chi'$ to $a + \sigma \ln L + c_1
L^{-0.4} + c_2 L^{-1}$ gives $\sigma = 2.78(9)$ ($L_{\rm min} = 8$) to be
compared with $\sigma = 2 - \eta + 1/\nu = 2.78(2)$, obtained by using our
estimates of $\nu$ and $\eta$.

In conclusion, we have characterized the scaling corrections to the asymptotic
critical behaviour in 3D Ising spin glass models.  We have shown that the
analytic $t$ dependence of the scaling fields gives rise to leading
$L^{-1/\nu}$ corrections ($1/\nu\approx 0.4$) in the FSS of some quantities.
In particular, these corrections appear in the derivative $\chi'$ at $T_c$.
This point has been apparently overlooked in earlier FSS studies. 
These analytic corrections may also be important in other glassy systems
in which $\nu$ is typically large. We have
estimated the leading {\em nonanalytic} scaling-correction exponent $\omega$
from the FSS of the quartic cumulants, obtaining $\omega=1.0(1)$.  Finally, we
have used these results to obtain accurate estimates of the critical exponents
$\nu$ and $\eta$. An analysis of the MC data that takes into account the
leading scaling corrections gives $\nu=2.53(8)$ and $\eta=-0.384(9)$.  Results
obtained by using different observables and different models are consistent.
This confirms that the bimodal Ising model belongs to a unique universality 
class, for any $p$ in the range $1 - p^* < p < p^*$, irrespective of bond 
dilution.


\end{document}